\documentclass[%
 reprint,
 amsmath,amssymb,
 aps,
prb,
floatfix,
]{revtex4-1}

\usepackage{graphicx}% Include figure files
\usepackage{epstopdf}

\usepackage{dcolumn}% Align table columns on decimal point
\usepackage{bm}% bold math
\usepackage[usenames,dvipsnames]{color}
\usepackage{soul}

\begin{document}

%\title{Cellular Dynamical Mean Field Theory Extension of the Typical Medium Theory for the Study of Anderson Localization}% Force line breaks with \\

\title{Real Space Quantum Cluster Formulation \\ for the Typical Medium Theory of Anderson Localization}% Force line breaks with \\

%\affiliation{Center for Computation & Technology, Louisiana State University, Baton Rouge, LA 70803, USA}

\author{Ka-Ming Tam}%
\email{phy.kaming@gmail.com}
\affiliation{Department of Physics and Astronomy, Louisiana State University, Baton Rouge, Louisiana 70803, USA}%

\author{Hanna Terletska}
\email{Hanna.Terletska@mtsu.edu}
\affiliation{Department of Physics and Astronomy, Middle Tennessee State University, Murfreesboro, TN 37132, USA}

\author{Tom Berlijn}
\affiliation{Center for Nanophase Materials Sciences, Oak Ridge National Laboratory, Oak Ridge, TN 37831, USA}

\author{Liviu Chioncel}
\affiliation{Theoretical Physics III, Center for Electronic Correlations and Magnetism, Institute of Physics, University of Augsburg, and Augsburg Center for Innovative Technologies, University of Augsburg, D-86135 Augsburg, Germany}

\author{Juana Moreno}
\affiliation{Department of Physics and Astronomy, Louisiana State University, Baton Rouge, Louisiana 70803, USA}%
\affiliation{Center for Computation \& Technology, Louisiana State University, Baton Rouge, LA 70803, USA}

\date{\today}% It is always \today, today,

\begin{abstract}
We develop a real space cluster extension of the typical medium theory (cluster-TMT) to study Anderson localization. By construction, the cluster-TMT approach is formally  equivalent  to  the  real  space cluster  extension  of  the  dynamical  mean  field  theory. Applying the developed method to the 3D Anderson model with a box disorder distribution, we demonstrate that cluster-TMT successfully captures the localization phenomena in all disorder regimes. As a function of the cluster size, our method obtains the correct critical disorder strength for the Anderson localization in 3D, and systematically recovers the  re-entrance behavior of the mobility edge. From a general perspective, our developed methodology offers the potential to study Anderson localization at surfaces within quantum embedding theory.

\end{abstract}
\maketitle

%\tableofcontents

\section{\label{sec:level1}Introduction}

The localization problem in disordered electronic systems was introduced in Anderson's seminal paper in the late fifties
and it still remains in the forefront of research in materials science and condensed matter physics~\cite{p_anderson_58}. To model disorder Anderson proposed a simplified model of electrons hopping between lattice sites being subject to static scattering processes on locally disordered centers. 
The stochastic character of the problem is encoded into the on-site energies
(disordered scattering centers) considered as random variables distributed according to a chosen probability distribution. The Green's function imaginary part, the local density of states (LDOS), turns out to be an important quantity which characterizes the disordered system. 
For example the LDOS is finite for extended states, while it is discrete for the localized states. A decade latter an approach based on the distribution of the site and energy dependent self-energies was formulated~\cite{Abou-Chacra_etal_1973}.
%$\Sigma_i(\omega)$, (with $i$ denoting the site index and $\omega$ the energy)  
This approach lead to a self-consistent equation for the self-energy, which can be exactly solved on a Cayley tree (Bethe lattice). Yet for general lattices only an approximate solution can be provided.    

Computations for substitutionally disordered three-dimensional materials with ordinary lattice structures are therefore difficult to perform within the framework of tight-binding models~\cite{p_anderson_58,Abou-Chacra_etal_1973}. 
Suitable modeling in such cases can be constructed based on effective medium theories. Among them single site effective medium methods, such as the coherent potential approximation (CPA)~\cite{p_soven_67,shiba71,b_velicky_68,s_kirpatrick_70,Onodera_Toyozawa_1968,d_taylor_67,Yonezawa_1968} and the typical medium theory (TMT)~\cite{v_dobrosavljevic_03}, proved to be simple and transparent theories that are able to capture important features of the disorder effects in electron systems.
Common to these two methods is the mapping of the lattice problem into the impurity placed in a self-consistently determined effective medium. In both methods the measured quantity is the disorder averaged Green's function, however in CPA the Green's function is linearly (algebraically) averaged, while in the TMT the geometric average of LDOS
%(imaginary part of the local Green's function)
is used. This difference in disorder averaging defines the average and the typical effective media, respectively. 

Unlike the algebraically averaged Green's function of the CPA effective medium, the geometric averaged LDOS, called the typical density of states (TDOS), drops to zero~\cite{v_dobrosavljevic_03,g_schubert_10,k_byczuk_05,Semmler_2010,Wortis_2011, Dobrosavljevic_int_2009,Dobrosavljevic_int_2013, Dobrosavljevic_int_2014, Dobrosavljevic_int_2015,Dobrosavljevic_review_2010,k_byczuk_09, k_byczuk_10}, at the Anderson transition. The geometrically averaged TDOS is an approximation to the most probable value in the distribution of the LDOS. At the Anderson transition, the system is not self-averaged, hence the distribution of the LDOS  is highly skewed with long tails~\cite{g_schubert_10, Fehske_2005}. Therefore, the average and most probable values of the LDOS will be very different close to the transition~\cite{g_schubert_10,m_janssen_94,m_janssen_98, Logan_Wolynes_1987}. Dobrosavljevic et al. ~\cite{v_dobrosavljevic_03} incorporated such statistical properties of the LDOS within the effective medium approach, called the  TMT. They showed that the TDOS successfully captures the main signatures of the Anderson localization transition, with the TDOS being an order parameter to detect the localized states.  In Refs. \onlinecite{c_ekuma_14b, c_ekuma_15b,Terletska_etal_2018} the momentum-space cluster extension of the TMT~\cite{Terletska_etal_2018}, the typical medium dynamical cluster approximation (TMDCA) has been developed. The TMDCA is the typical medium extension of the Dynamical Cluster Approximation (DCA) ~\cite{m_jarrell_01a,m_jarrell_01c}, a momentum-space cluster extension of the CPA. The TMDCA overcomes the shortcomings of the local single site TMT, and accurately predicts the critical disorder strength of the Anderson localization transition in a single-band Anderson model. For model Hamiltonian systems, the TMDCA has been applied to non-interacting and weakly interacting disordered three dimensional systems ~\cite{c_ekuma_15b,c_ekuma_14b,s_sen_16a, h_terletska_17}, systems with off-diagonal disorder~\cite{h_terletska_14a}, phonon localization~\cite{Mondal_etal_2019,Mondal_2020} and multi-orbital models~\cite{Zhang_etal_2015}. Some of the methods inspired by the typical medium theories have been combined with the first-principles calculations~\cite{Zhang_etal_2016,Zhang_etal_2018,Ostlin_etal_2020}. 

Complementary to the momentum space cluster methods, described above, techniques using embedding in real space provide an interesting alternative. This constitutes the aim of the present work. We have previously formulated the embedding into the effective typical medium which allows to address the Anderson localization transition in the framework of locally self-consistent approach~\cite{Zhang_etal_2019}. Besides, the locally self-consistent formulation it opens the possibility to formulate linear scaling methods. Unlike the previous typical medium cluster extensions of TMT, formulated in the momentum  space (TMDCA)~\cite{Terletska_etal_2018,Terletska_etal_2021}, or in a mixed representation (locally self-consistent approach)~\cite{Zhang_etal_2019,Tam_etal_2021}, here we propose an exclusively real space cluster extension of TMT (cluster-TMT). This construction is formally equivalent to the real space cluster extension of the dynamical mean field theory (DMFT)~\cite{a_georges_93,Biroli_Kotliar_2002,Biroli_etal_2004,g_kotliar_01,Lichtenstein_Katsnelson_2000}. Characteristic to the present cluster extension of the TMT  is the matrix form of the TDOS with diagonal and off-diagonal elements. Applying our real space cluster-TMT to 3D Anderson model with a box disorder distribution, we find that cluster extensions of TMT are necessary to properly capture the non-local effects in the Anderson transition. Quantitatively our results are in a good agreement with the existing data in the literature, in particular, we find that the converged cluster value of $W_c\approx 17.05$ is superior to the value of $13.4$ provided by single site TMT calculations.
Unlike the single site TMT, the present real-space cluster computation captures the re-entrance behavior driven by non-local multiple scattering effects which are missing in local approximation \cite{v_dobrosavljevic_03, Kramer_MacKinnon_1993,b_bulka_85,Fehske_2005}.
%\cite{Kramer_MacKinnon_1993,b_bulka_85,b_bulka_87,deQueiroz_2001}.
Just like the TMDCA, the real space cluster-TMT allows for a computationally efficient treatment of the non-local effects in Anderson localization. In addition however, the cluster TMT opens the door to open boundary conditions, which offers the possibility to study the localization of surface states. One potential application of this capability would be the search for a materials realization of the topological Anderson insulator via first principles calculations ~\cite{Li_2009_prl}.

%By a suitable choice of the  averaging of the off-diagonal elements ... mobility edge...
This paper is organized as follows: In Sec.~\ref{sec:II}, we present the Anderson model. In Sec.~\ref{sec:III}, we first briefly review the algorithm of the single site TMT and discuss the algorithm for the real-space cluster extension of the TMT. In Sec.~\ref{sec:IV}, we present the results obtained with our cluster-TMT to the 3D Anderson model with the box disorder distribution. 
We conclude in Sec.~\ref{sec:V} and discuss possible future developments.

\section{Model}
\label{sec:II}
Anderson proposed that non-interacting electrons on site-disordered lattices may localize because of the destructive interference of wave functions~\cite{p_anderson_58}. Subsequent theoretical and numerical studies supports the picture that in three dimensions and for sufficient disorder single particle wave functions are localized at band edges and decay exponentially on the scale of the localization length~\cite{p_lee_85}.

%\subsection{Anderson Model}
%To study such disorder effects, we consider the standard Anderson model
The Anderson model Hamiltonian has the form:
\begin{equation}
    H = -t \sum_{<{i,j}>,\sigma} (c^{\dagger}_{i\sigma} c_{j\sigma} + H.c.) + \sum_{i\sigma} V_{i} n_{i\sigma},
\end{equation}
where $c^{\dagger}_{i\sigma}$ and $c_{i\sigma}$ are the creation and annihilation operators for electrons at site $i$ with spin $\sigma$. $n_{i,\sigma}$ is the number operator for site $i$ of spin $\sigma$; $t$ is the hopping energy between nearest neighbors. 
%We assume the lattice is at dimension, $D=3$, of a simple cubic lattice. 
We consider a 3D simple cubic lattice.
We set $t=1$ to serve as the energy scale. The local random disorder is given by $V_{i}$. Here we  consider a so-called box disorder with $P(V_i) = \frac{1}{W}\Theta(|W-V_i|)$. This allows the disorder strength to be characterized by $W$. 
Other distributions are also considered in the literature, some common ones included bi-modal, Gaussian, and Lorentzian distributions \cite{Selvan08,c_ekuma_15b}.
%For the calculations in this paper, a finite number, $N_r$ of random realizations is used for averaging. 

The Anderson model has been the focus of numerous studies of the disorder-induced electron localization. 
%from the original paper by Anderson. \cite{p_anderson_55} 
Highly accurate numerical calculations based on the transfer matrix method and multifractal analysis have been used to study the model extensively, especially for the zero energy \cite{b_bulka_85,b_bulka_87,Kramer_MacKinnon_1993,b_kramer_87,b_kramer_10,a_rodriguez_10,a_rodriguez_11,k_slevin_01,k_slevin_14,k_slevin_99,Chang_etal_1990,MacKinnon_Kramer_1983}.

Relatively few studies have been devoted to energy away from zero. A prominent feature at higher energy is the re-entrance from a metal to an insulator to a metal, as the disorder strength increases~\cite{deQueiroz_2001,b_bulka_85,b_kramer_87,h_grussbach_95}. 
A heuristic argument for the nature of the re-entrance  behavior is based on the tunneling mechanism for energies beyond the bandwidth of the hopping model. The width of the density of states increases as the disorder increases, though the states are localized.  
At sufficiently large disorder the localized density of states is large enough to allow tunneling. The tunneling could become sufficiently long range that the localized states become extended thus the insulator becomes a metal. This explains the lower transition in the re-entrance. Further increasing the disorder strength, the localized state will be more sparse in energy and tunneling becomes less likely to happen and insulating state resumes. 

The above argument depends on the distribution of disorder, the tunneling effect is maximised when the localized states are close in energy. A bounded random distribution is favored as compared to other distributions which are more widely spread over a range of energy, such as the Lorentzian distribution. 
The tunneling argument can only be supported in a system with multiple sites. For example the TMT, which is a single site approximation, does not capture the  re-entrance behavior. Thus, the capability of describing the re-entrance can serve as a good test for our real space cluster-TMT.
%the cluster method. 

\section{The Real Space Quantum Cluster Extension of TMT }
\label{sec:III}
\subsection{Typical Medium Theory: TMT}

To set the stage for the discussion of the real-space cluster extension of the TMT, here we briefly review the main steps of the TMT analysis. The TMT can be considered as a typical medium generalization of the CPA~\cite{p_soven_67,shiba71,b_velicky_68,s_kirpatrick_70,Onodera_Toyozawa_1968,d_taylor_67,Yonezawa_1968}. Similar, to the CPA, the TMT employs the mapping of the original lattice problem into the impurity placed in a self-consistently determined effective medium. However, in the TMT, the typical (geometrically averaged over disorder) local density of states is used to construct the mean field bath for the effective impurity problem. 

\begin{figure}[t!]
    \centering
\includegraphics[width=0.5\textwidth]{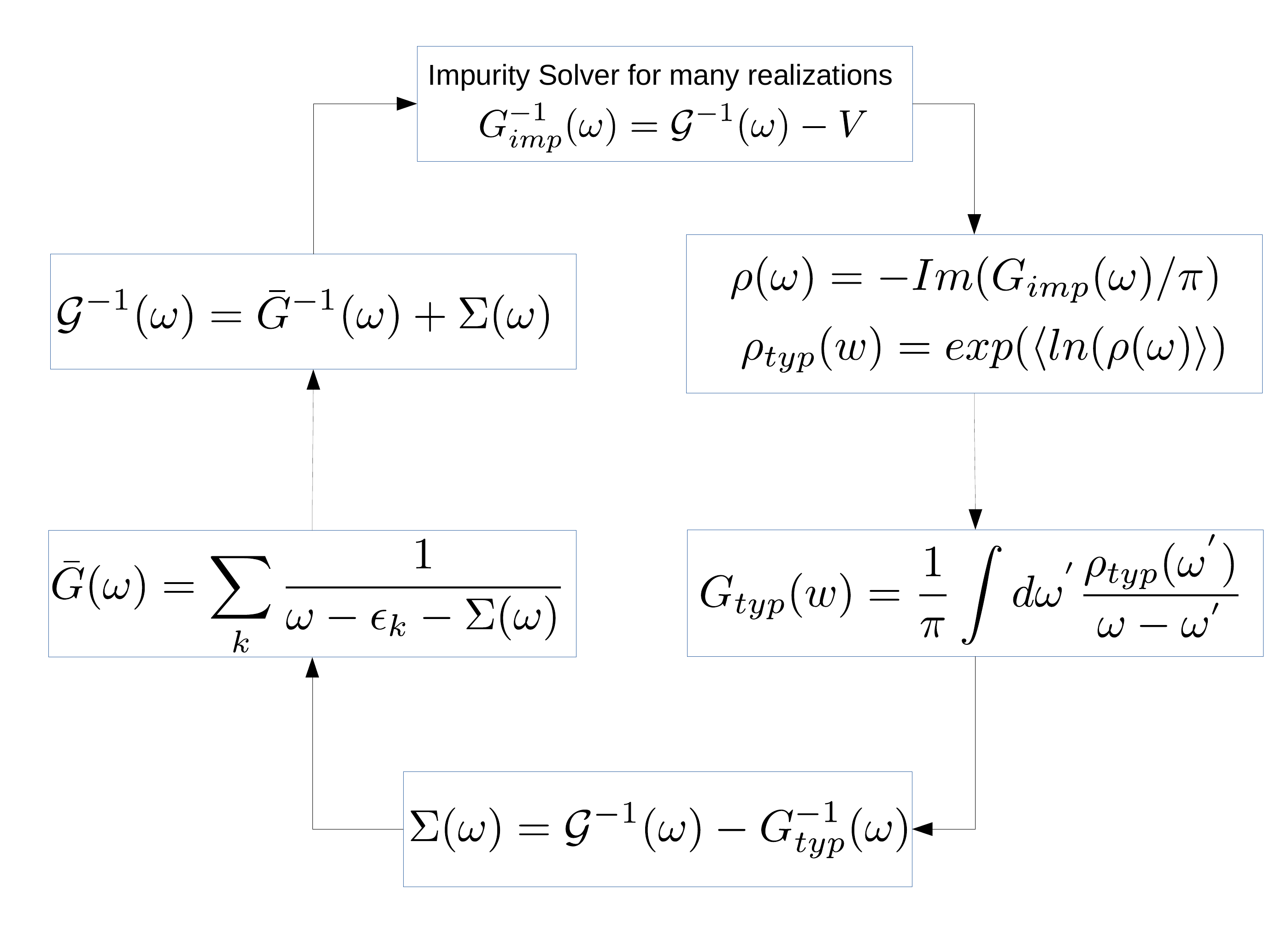}
    \caption{Numerical algorithm for the typical medium theory. % 1. Impurity problem is solved. For the non-interacting problem, it is simply given by taking the inverses $G_{c}=(\mathcal{G}^{-1}+V)^{-1}$. Repeat it for many 
   % random realizations, $V$. 2. Calculating the typical density of state from the set of impurity Green's function. 3. The typical Green's function is calculated via the Hilbert transform of the typical density of state. 4. The self energy is obtained by the Dyson equation. 5. The coarse grained lattice Green's function is obtained by averaging over lattice Green's function. 6. The Dyson equation is used to extract the bath Green's function for the impurity problem. 7. Repeat from step 1 until the self energy is converged. 
   }
    \label{fig:tmt_algorithm}
\end{figure}

The numerical algorithm for the TMT procedure is shown in Fig.~\ref{fig:tmt_algorithm}. First, the guess for the effective medium self-energy $\Sigma(\omega)$ is made, usually zero. Then, the local (coarse-grained) lattice Green's function is calculated as $\bar{G}(\omega) = \frac{1}{N}\sum_{k} \frac{1}{\omega-\epsilon_{k}-\Sigma(\omega)}$.
Using the Dyson's equation, we then obtain the impurity-excluded Green's function (bath Green's function) $\mathcal{G}^{-1}(\omega)= \bar{G}^{-1}(\omega) + \Sigma(\omega)$. 

The next step is to solve the impurity problem. For each randomly chosen disorder configuration $V$, we calculate the impurity Green's function $G_{imp}(\omega,V)=(\mathcal{G}^{-1}(\omega)-V)^{-1}$. %where $V(i,j)=V_{i}\delta_{i,j}$ is one of the realizations of local random potential draw from a desired distribution. 
From this quantity, we obtain the typical (geometrically averaged density of states) $\rho_{typ}(\omega)$, which is constructed as $\rho_{typ}(\omega) = e^{\langle ln(\rho(\omega,V))\rangle}$. Here, $\rho(\omega, V)=-\frac{1}{\pi}\Im G_{imp}(\omega, V)$, and $\langle ... \rangle$ stands for the disorder averaging. In general, the geometrical average is not equivalent to the typical value. Numerical studies have shown that near the localization transition the local density of states is log-normal distributed \cite{g_schubert_10}. For log-normal distribution the geometrical average is the same as typical value. 

The output of the TMT impurity solver is the typical Green's function which is obtained using the Hilbert transform $G_{typ}(w) = \frac{1}{\pi}\int d\omega^{'} \frac{\rho_{typ}(\omega^{'})}{\omega-\omega^{'}}$. This step is the only difference between the CPA and the TMT self-consistency loop. I.e., in the CPA, instead of the typical, the algebraically average DOS is calculated $\rho_{ave}=\langle\rho(\omega,V)\rangle$, with the average Green's function $G_{ave}(w) = \frac{1}{\pi}\int d\omega^{'} \frac{\rho_{ave}(\omega^{'})}{\omega-\omega^{'}}$ being the output of the CPA impurity solver. Note, that for the CPA case, one can just do the disorder averaging over Green's function without the Hilbert transform of the average density.

Finally, the TMT self-consistency loop is closed by getting a new estimate of the self-energy $\Sigma(\omega) = \mathcal{G}^{-1}(\omega) - G_{typ}^{-1}(\omega)$, which is then used to calculate the coarse-grained local lattice Green's function. The whole procedure then repeats, until convergence is reached at which the impurity and the local lattice Green's function are equal with the desired accuracy.

\subsection{Real Space cluster-TMT}

To properly capture the multiple impurities scattering effects in the disorder-driven Anderson localization, the cluster extension of the TMT is needed. Here we present the real-space cluster extension of the TMT. Such real space variant of the cluster extension of the TMT is formally equivalent to the cluster DMFT solver, which has been extensively used in strongly-correlated electron systems to study beyond DMFT non-local effects. Here we use the cluster DMFT approach to the disordered non interacting systems, and use it as a tool to capture spacial non-local correlations beyond the TMT.

In the real space cluster-TMT, the infinite lattice is tiled with identical clusters of size $N_c$ in real space~\cite{Senchal_2010}. In such construction, the scattering of electrons on impurities within a cluster
is treated exactly, while those outside the cluster are replaced by the non-disordered effective medium (bath) that is determined self-consistently.  There is no implicit assumption that the transnational invariance is obeyed within the cluster. Therefore the Green's function of the cluster is represented by an $N_c \times N_c$ matrix, which we denote as $\hat{G}_c(\omega)$. For the same reason, the self-energy, the bath Green's function are also represented in term of matrices. 

\begin{figure}[t!]
    \centering
    \includegraphics[width=0.5\textwidth]{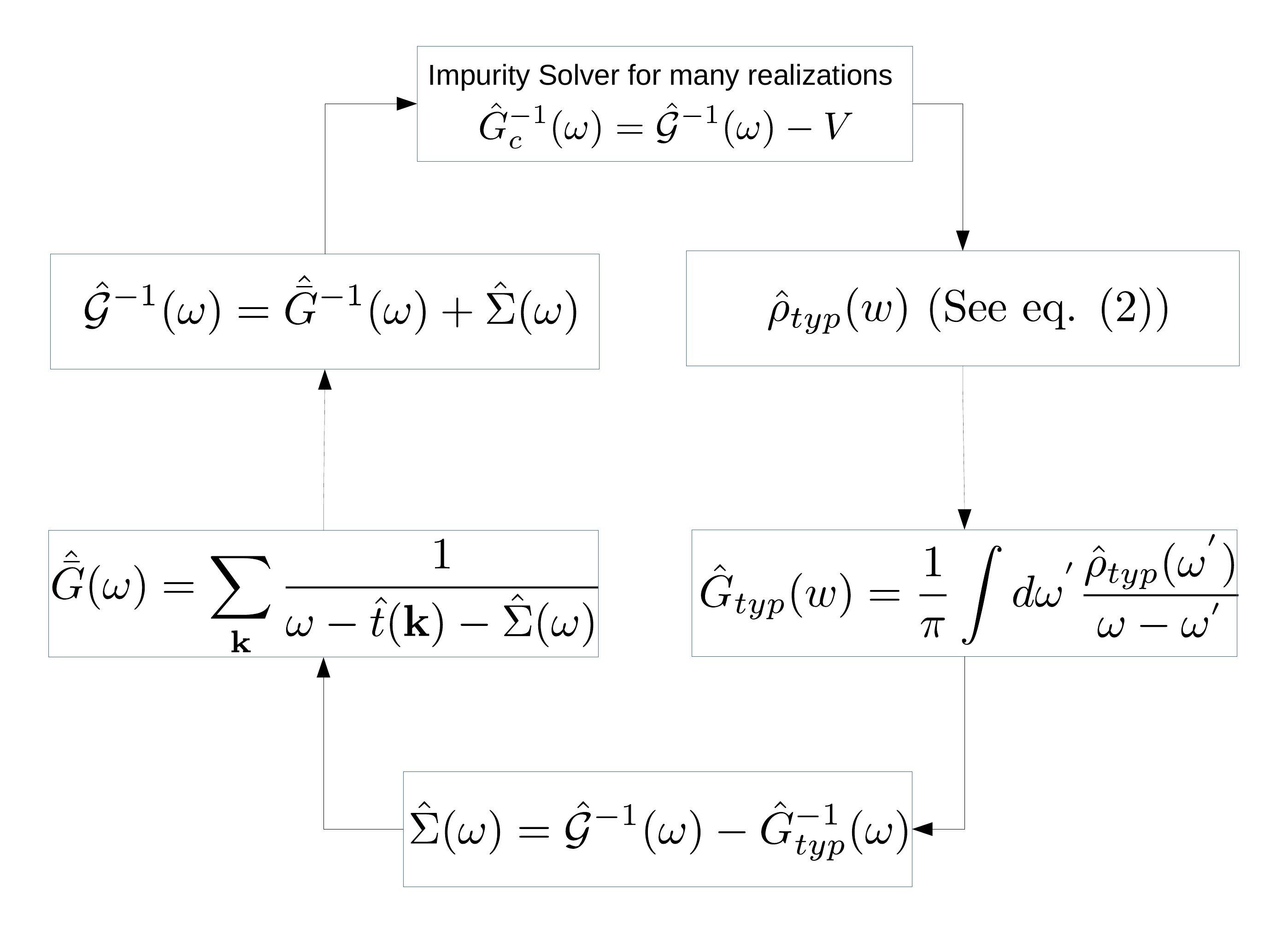}
    \caption{The self-consistency algorithm for the real space cluster-TMT formalism.
    %typical medium theory cellular dynamical mean field theory. 1. Impurity problem is solved. Repeat it for many 
    %random realizations, $V$. 2. Calculating the typical density of state from the set of impurity Green's function by eq. \ref{eqn:ansatz}. 3. The matrix elements of the typical Green's function are calculated via the Hilbert transform of the typical density of state one by one. 
    %4. The self energy is obtained by the Dyson equation. 5. The coarse grained lattice Green's function is obtained by averaging over lattice Green's function within the reduced Brillouin zone. 6. The Dyson equation is used to extract the bath Green's function for the impurity problem. 7. Repeat from step 1 until the self energy is converged. 
    }
    \label{fig:tmt_cdmft_algorithm}
\end{figure}

The self-consistency procedure for our real space cluster-TMT is shown in Fig.~\ref{fig:tmt_cdmft_algorithm}. First, we start with the guess of the self-energy matrix $\hat{\Sigma}(\omega)$ (usually zero), and calculate the lattice Green's function projected onto the cluster space $\hat{\bar{G}}(\omega)=\frac{N_c}{N}\sum_{\mathbf{k} \in R.B.Z.}[\omega-\hat{t}(\mathbf{k})-\hat{\Sigma}(\omega)]^{-1}$, R.B.Z. stands for the reduced Brillouin Zone of the cluster with $\frac{2\pi}{L_c}<k_x,k_y,k_z<\frac{2\pi}{L_c}$, $\hat{t}(\mathbf{k})$ is the dispersion of the lattice model expressed as a partial Fourier transform over the reduced Brillouin zone (with  $\hat{t}_{\mathbf{r},\mathbf{r^{'}}}(\mathbf{k}) \equiv \sum_{\mathbf{R}} exp(i\mathbf{k}\cdot(\mathbf{R}+\mathbf{r} - \mathbf{r^{'}}))t_{\mathbf{r},\mathbf{r^{'}}+\mathbf{R}}$, where $\mathbf{R}$ is the vector for the location of the super-cells; $\mathbf{r}$ and $\mathbf{r^{'}}$ are the vectors for the location of the sites within a super-cell.
 \cite{Senchal_2010}.

Next, using the Dyson's equation,  we calculate the bath Green's function matrix, $\hat{\mathcal{G}}^{-1}(\omega)= \hat{\bar{G}}^{-1}(\omega) + \hat{\Sigma}(\omega)$, which is used to construct the cluster problem. Notice that unlike in the momentum-space cluster extension of TMT, here the bath is not diagonal. Then, for each disorder configuration $V$, we calculate the cluster Green's function by solving the matrix equation $G_{c}(\omega,i,j)=(\mathcal{G}^{-1}(\omega,i,j)-V(i,j)\delta_{ij})^{-1}$. 

The key to incorporate the typical medium into the analysis is to connect the Green's function matrix to the typical density of states. For this, we generalize the procedure we used for the  multi-orbital problem of the TMDCA \cite{y_zhang_15a}, and define the typical density of states matrix in a similar way: 
    
\begin{eqnarray}\label{eqn:ansatz}
    &&
    \hat{\rho}_{typ}(\omega)\equiv \\ \nonumber %F(G_{c,ij}^{r}) \\ \nonumber
   &&  
\left(
\begin{array}{cccc} 
e^{\left\langle |\rho_{11}(\omega)|\right\rangle} \frac{<\rho_{11}>}{<|\rho_{11}|>}   	& \cdots	 & e^{\left\langle |\rho_{1N_c}(\omega)|\right\rangle} \frac{<\rho_{1N_c}>}{<|\rho_{1N_c}|>}	 \\
        .		 	&	 .	 &     .		 \\
	.		 	&	 .	 &     .		 \\
	.		 	&	 .	 &     .		 \\
e^{\left\langle|\rho_{N_c1}(\omega)|\right\rangle} \frac{<\rho_{N_c1}>}{<|\rho_{N_c1}|>}		 	& \cdots	 & e^{\left\langle |\rho_{N_cN_c}(\omega)|\right\rangle} \frac{<\rho_{N_cN_c}>}{<|\rho_{N_cN_c}|>}  \
\end{array} 
\right), \\ \nonumber
\end{eqnarray}
Here the diagonal entries will be just equal to $e^{\left\langle \rho_{ii}(\omega)\right\rangle}$, because $\rho_{ii}>0$ is always positive definite; %The density of state is calculated from the Green's function. 
 $\rho_{ii} = -\frac{1}{\pi} \Im[G_{ii}(\omega)]$; and for the off-diagonal terms $\rho_{ij} = \frac{i}{2\pi} \Im[G_{ij}(\omega)-G_{ji}(\omega)]$ \cite{Kraberger_etal_2017}.
 %Having the typical density of state matrix $\hat{\rho}_{typ}$, we can obtain the typical Green's function matrix by performing the Hilbert transform of each matrix elements individually, $G_{typ,ij}(w) = \frac{1}{\pi}\int d\omega^{'} \frac{\rho_{typ,ij}(\omega^{'})}{\omega-\omega^{'}}$. 
 
 Notice that the real space cluster extension of the CPA, with the average effective medium, can be obtained by replacing the typical DOS with the linearly average DOS in the above Eq.~\ref{eqn:ansatz}, i.e.
 
\begin{eqnarray}\label{eqn:ansatz-ave}
    &&\hat{\rho}_{ave}(\omega)\equiv   \left(\begin{array}{cccc} \langle \rho_{11}(\omega)\rangle    	& \cdots	 & \langle \rho_{1N_c}(\omega)\rangle	 \\
        .		 	&	 .	 &     .		 \\
	.		 	&	 .	 &     .		 \\
	.		 	&	 .	 &     .		 \\
\langle \rho_{N_c1}(\omega)\rangle		 	& \cdots	 & \langle \rho_{N_c N_c}(\omega)\rangle  \\

\end{array} \right), \\ \nonumber
\end{eqnarray}

The $\hat{\rho}_{typ}(\omega)$ of Eq.~\ref{eqn:ansatz} possesses the following properties: 1) for $N_c=1$, it reduces to the local TMT with $\rho_{typ}(\omega)=e^{\langle\rho(w)\rangle}$; 2) At low disorder strength $W \ll W_c$, we observe numerically that $<\ln \rho(\omega)>\approx \ln<\rho(\omega)>$, i.e., the typical density of states (DOS) reduces to the average DOS calculated using algebraic averaging over disorder, with $\rho_{typ}\rightarrow \rho_{ave}(\omega)$. Hence, in this regime the typical DOS obtained with the cluster-TMT is expected to be close in magnitude to the one obtained with the real-space cluster-CPA with averaged effective medium. Such real space cluster extension of CPA is different from other existing cluster extensions, including the DCA~\cite{m_jarrell_01a,m_jarrell_01c} and non-local CPA~\cite{Rowlands_2006}. The difference is that in the real space cluster-CPA, all the quantities are matrices in the real space, and the coarse-graining step for $\bar{G}$ uses a projected lattice dispersion on to the  real space cluster  space.

In the next step of the cluster-TMT self-consistency loop, we must calculate the cluster typical Green's function $\hat{G}_{typ}$ ($\hat{G}_{ave}$ for the cluster-CPA, respectively) using the Hilbert transform. The Hilbert transform is performed for each matrix element individually, $G_{typ,ij}(w) = \frac{1}{\pi}\int d\omega^{'} \frac{\rho_{typ,ij}(\omega^{'})}{\omega-\omega^{'}}$.

Next, using the Dyson's equation, we get the updated self-energy $\hat{\Sigma}(\omega) = \hat{\mathcal{G}}^{-1}(\omega) - \hat{G}_{typ}^{-1}(\omega)$, which is then used to calculate the coarse-grained lattice Green's functions matrix $\hat{\bar{G}}$. The whole procedure then repeats, until convergence is reached with the desired accuracy.

\section{Results}
\label{sec:IV}
%%%%%%%%%%%%%%%%%%%%%%%%%

\begin{figure}[t!]
    \centering
    \includegraphics[width=0.45\textwidth]{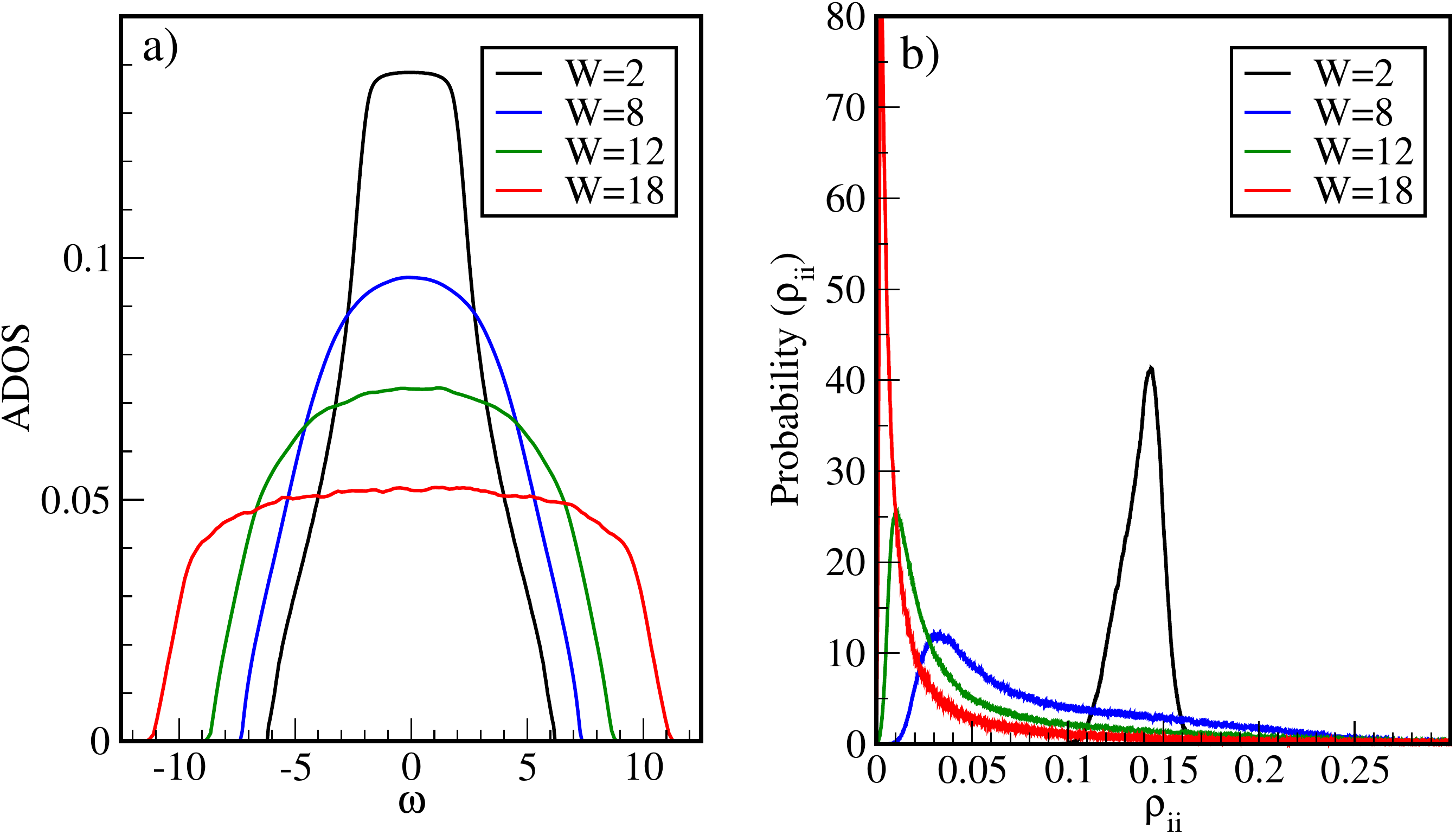}
\caption{'a) The ADOS calculated for $N_c=3^3$ at several disorder strengths $W=2, 8, 12, 18$; b) The probability distribution function of the local density of states $\rho_{ii}$ for several values of disorder strengths, $W=2, 8, 12, 18$.}
  \label{fig:histogram_rho}
\end{figure}

We start the discussion of our results for $3$D Anderson model (for a box disorder distribution) by first showing in Fig.~\ref{fig:histogram_rho} (a-panel) the $N_c=3^3$ cluster average DOS (ADOS=$\frac{1}{N_c}\Sigma_i \Im\hat{G}_c(ii,\omega)$) obtained using the average effective medium (constructed from Eq.~\ref{eqn:ansatz-ave} ) in the cluster self-consistency loop. These results correspond to the real-space cluster extension of the CPA. The data show that as disorder strength $W$ increases, the ADOS broadens and gets smaller, but does not go through significant qualitative changes when the metal-insulator transition is approached. 

To demonstrate why the ADOS fails to describe the Anderson transition, in Fig.~\ref{fig:histogram_rho} (a-panel), we consider the probability distribution of the local density of states. At small disorder $W=2$, the distribution of the LDOS is Gaussian-like. However, as disorder strength increases, the probability distribution becomes skewed with long tails (indicating that the system is not self-averaging), and peaks close to zero values at even larger disorder strength ($W=18$). 
Such skewness in the distribution functions at larger disorder W implies that the measured average and the most probable (typical) values of the DOS will differ significantly, and hence the numerical algorithms that employ the globally averaged Green's function in the self-consistency loop (e.g., the CPA and the DCA) will fail to describe the Anderson transition. 

\begin{figure}[t]
    \centering
    \includegraphics[width=0.45\textwidth]{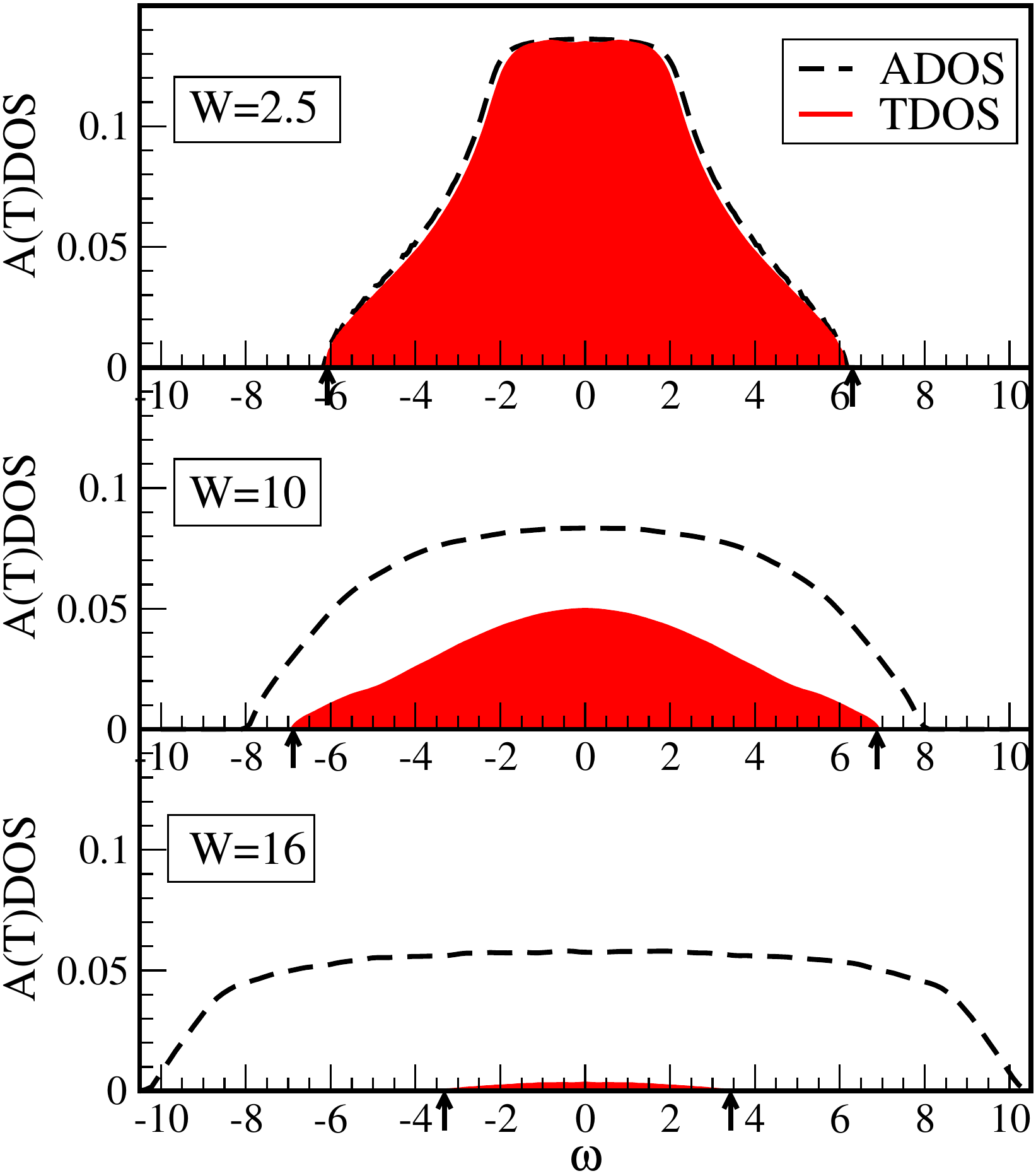}
\caption{ Evolution of the ADOS (dash lines) and the TDOS (shaded areas) as function of frequency $\omega$ at different disorder strengths $W=2.5, 10, 16$, calculated using the $N_c=3^3$. The approximate positions of the mobility edge boundaries are marked by vertical arrows.}
  \label{fig:tdos_ados_W}
\end{figure}

These results clearly demonstrate that the typical medium treatment is required to capture the non self-averaging behavior through the Anderson transition. To show this, in Fig.~\ref{fig:tdos_ados_W}, we 
%We start the discussion of our results for $3$D Anderson model (for a box disorder distribution) by
compare the data for the energy resolved ADOS and the TDOS calculated for a clsuter of $N_c=3^3$ sites (Fig.~\ref{fig:tdos_ados_W}). 
%The ADOS$(\omega)$ is obtained using the CDMFT scheme with the arithmetically averaged over disorder effective medium, while 
The TDOS$(\omega)=exp(\frac{1}{N_c}\Sigma_i log(\Im \bar{G}_c(ii,w)))$ is obtained from the present real-space cluster-TMT procedure which employs the geometric averaging in the self-consistency loop. 
%{(\color{red} Please check the change in the sentence above)}
%The A(T)DOS$(\omega)=\frac{1}{N_c}\sum_iG_c(ii,\omega)$ are obtained from the converged average (typical) cluster Green's function, respectively. 
At weak disorder strength $(W=2.5)$, as expected from our analytical arguments, both ADOS and TDOS are practically the same, indicating that at $W\ll W_c$ the real space cluster-TMT reduces to cluster-CPA scheme. As disorder strength increases, the ADOS and TDOS behave very differently. While the ADOS$(\omega)$ broadens and remains finite, the TDOS$(\omega)$ gets continuously suppressed ($W=10$) and vanishes at even larger disorder strength ($W=16$). Such vanishing of the TDOS at strong disorder $W$ indicates that geometrically average DOS can be used as an order parameter for the Anderson localized states. 

Notice, that below the Anderson transition, for $W \ll W_c$, localization of states starts at the band tails. This is indicated by vanishing TDOS$(\omega)$ and a finite ADOS$(\omega)$ at higher frequencies $\omega$. The mobility edge (shown by arrows), i.e. the energy which separates the extended (with a finite TDOS) and the localized states (with zero TDOS) follows the expected re-entrance trajectory~\cite{c_ekuma_14b}: the mobility edge first expands beyond zero disorder edge boundary, and then retracts at larger disorder strength.

%%%%%%%%%%%%%%%%%%%%%%%%%%%%%%%%%%%%%%%%%%%
\begin{figure}[t!]
    \centering
    \includegraphics[width=0.45\textwidth]{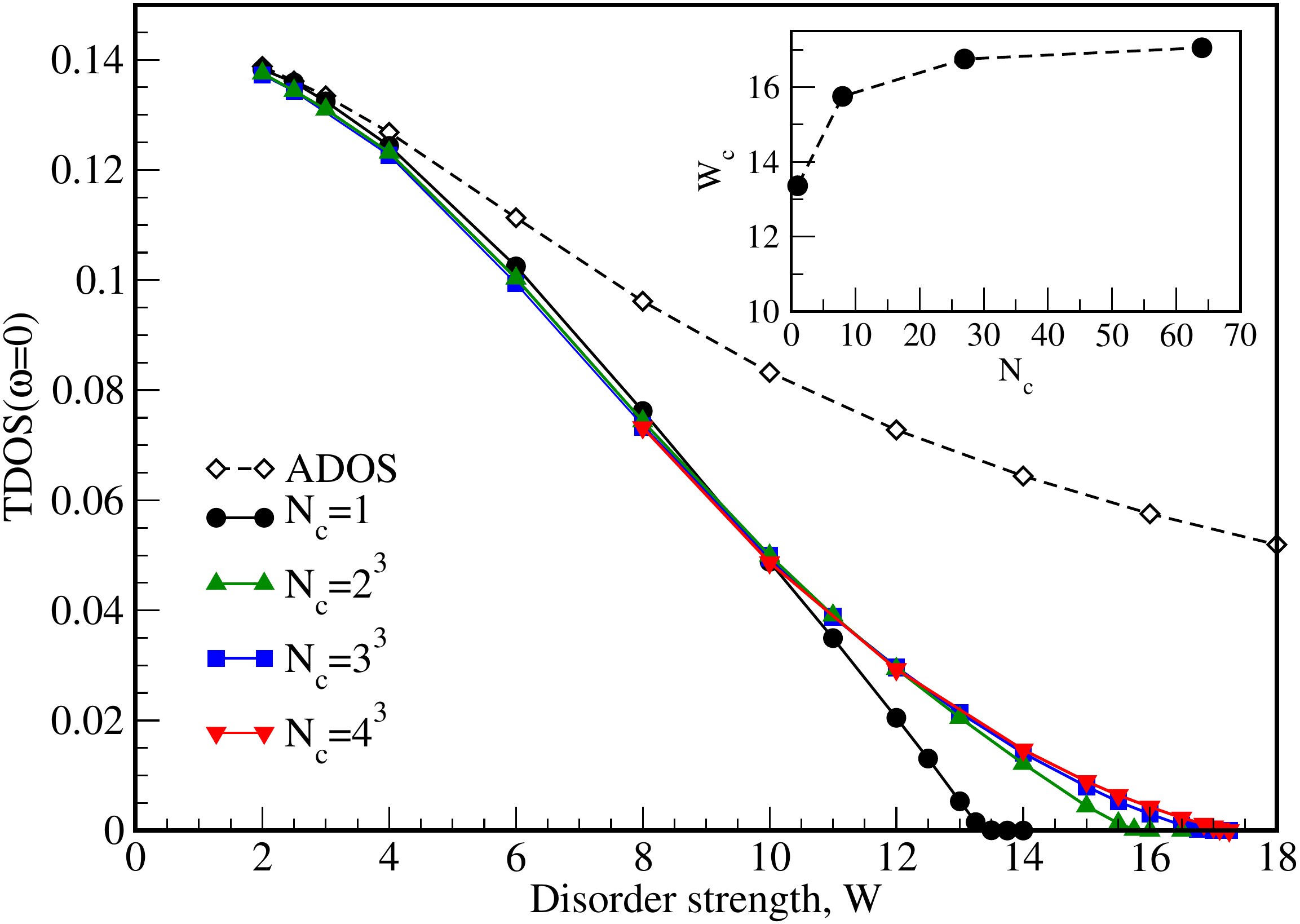}
\caption{The typical density of states (solid lines) at the band center, $TDOS(\omega=0)$, as a function of disorder strength $W$ calculated for different cluster sizes $N_c=1, 2^3, 3^3, 4^3$. The ADOS$(\omega=0)$ as a function of disorder strength $W$ is obtained for $N_c=4^3$ (a dash line). Inset: the cluster size $N_c$ dependence of the critical disorder strength $W_c$ determined from the vanishing TDOS$(\omega=0)$. 
}
\label{fig:tdos_zero}
\end{figure}
%%%%%%%%%%%%%%%%%%%%%%%%%%%%%%%%%%%%%%%%%%%%%
Next, in Fig.~\ref{fig:tdos_zero}, we consider the evolution of the critical disorder strength $W_c$ for the Anderson transition as a function of the cluster size $N_c$. The critical disorder $W_c$ is extracted from the vanishing TDOS at the band center (TDOS$(\omega=0)$). In Fig.~\ref{fig:tdos_zero} we plot TDOS$(\omega=0)$ as a function of disorder strength $W$ for several cluster sizes $N_c=1, 2^3, 3^3, 4^3$. For $N_c=1$ (the local TMT case), the critical disorder $W_c\approx 13.4$. Since TMT is a mean field theory, it is expected that the critical disorder strength is underestimated and thus it is lower than the exact value. As the cluster size $N_c$ increases, more spatial fluctuations are taken into account, which improves the value of $W_c$. With increasing $N_c$, the $W_c$ converges quickly to $W_c \approx 17.05$ (see inset of Fig.~\ref{fig:tdos_zero}), which is in good agreement with the values of $W_c$ reported in the literature~\cite{MacKinnon_Kramer_1983}. Also notice that unlike the TDOS, the ADOS$(w=0)$ (shown by the dashed line in Fig.~\ref{fig:tdos_zero}) remains finite as the disorder strength $W$ increases, indicating that it can not be used as an order parameter for the Anderson transition, and hence the typical medium treatment is needed.

%%%%%%%%%%%%%%%%%%%%%%%%%%%%%%%%%%%%%%%%%%%%%
\begin{figure}
    \centering
    \includegraphics[width=0.45\textwidth]{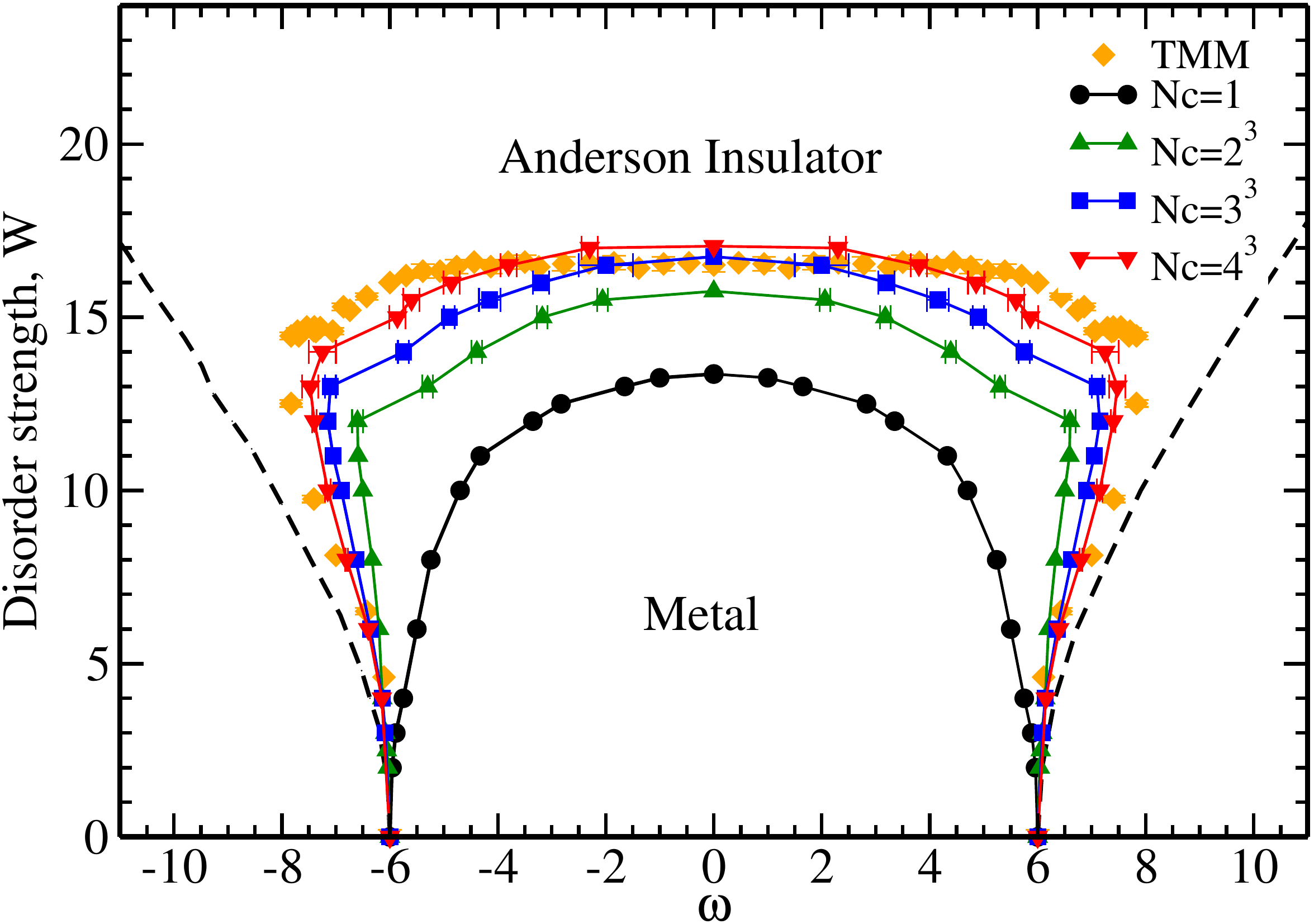}
\caption{Disorder strength $W$ vs frequency $\omega$ phase diagram of $3$D Anderson model obtained from cluster-TMT calculations. The mobility edge boundaries (solid lines) are obtained for $N_c=1, 2^3, 3^3, 4^3$ cluster sizes. Dashed line mark the band edges obtained from the ADOS$(\omega)$. The transfer matrix method (TMM) mobility edge boundaries are taken from Ref. \onlinecite{c_ekuma_14b}.}
\label{fig:PD}
\end{figure}

Finally, in Fig.~\ref{fig:PD}, we present the disorder strength $W$ vs. frequency $\omega$ phase diagram. Here we plot the cluster size $N_c$ dependence of the mobility edge boundaries at different disorder strengths $W$ obtained by our real space cluster-TMT formalism.
%which at a given disorder strength $W$ is given by the frequencies at which TDOS$(\omega)=0$. 
In addition, we also show the band edges, which are defined by frequencies at which ADOS$(w)=0$. As we discussed above a signature of the cluster mean field theory is the re-entrance at high energy. At $N_c>1$, the mobility edge boundaries first expand and then retract back with increasing $W$. As seen from the Fig.~\ref{fig:PD}, such re-entrance behavior is missing in the single site($N_c=1$) TMT case, and is recovered for $Nc>1$ results. This indicates that non-local spacial correlations and multiple-scattering effects in the Anderson transition are important, and capturing such effects requires the usage of a finite cluster methods. To benchmark our results even further, we also present the mobility edge trajectories obtained from the highly accurate transfer matrix method (TMM) ~\cite{Terletska_etal_2018}. For $N_c=4^3$, the cluster-TMT results are already rather close to that of the TMM. These results demonstrate that our cluster-TMT method can be used to successfully describe the electron localization in 3D Anderson model.

\section{Conclusion}
\label{sec:V}

We develop a real space quantum cluster theory based on the idea of the typical medium theory for random disorder systems. Unlike the coherent potential approximation with the average effective medium, the typical medium theory captures the localization transition by considering the geometrically averaged local density of states to construct an effective medium. However, being a single site theory, the TMT underestimates the critical disorder strength of the transition, and misses re-entrance behavior which is due to the combined effects from the multiple sites.
%the re-entrance effect due to the tunneling is the prime example for the Anderson model. 
Recent studies based on the dynamical cluster approximation already confirmed that such non-local effects can be captured by considering the  momentum-space clusters extension of TMT~ \cite{Terletska_etal_2018}.

In this paper, we construct the real space variant of the cluster-TMT. This method by construction is similar to another popular cluster method the cellular dynamical mean field theory effectively used for strongly interacting electron systems. Here we adopt such a real space cluster approach to disordered systems. Applying our real-space cluster-TMT approach to the 3D Anderson model with a box distribution, we demonstrate that cluster-TMT presents a successful self-consistent numerical approach for Anderson localization. Performing $N_c$ cluster-size analysis, we demonstrate the importance of the non-local spacial effects to properly describe the Anderson localization physics. Quantitatively our results are in a good agreement with the existing data in the literature, in particular, we find that the converged cluster value of $Wc^{cluster-TMT} \approx 17.05$ is superior to the value of $W^{TMT}\approx 13.4$ provided by a single site TMT calculations. Unlike the single site approach, the present real-space cluster-TMT  captures  the  re-entrance  behavior and reproduce correctly the phase diagrams of the 3D Anderson model. The method, in principle, can also be used to calculate two particle quantities \cite{y_zhang_17}. 
Furthermore, while the cluster TMT in this study has been restricted to periodic boundary conditions, the same methodology can be used to simulate Anderson localization in surfaces. This will be relevant for example for unraveling the role of disorder in topological materials~\cite{Li_2009_prl,Bitan_2018_prx}. Another interesting topic is to incorporate it with the multiple scattering theory \cite{h_terletska_17} and locally self-consistent multiple scattering method \cite{Zhang_etal_2019} for the study of materials with random disorder.

\section{Acknowledgement}
The authors would like to thank V. Dobrosavljevic and S. Iskakov for useful comments and discussions.

This manuscript is based upon work supported by the U.S. Department of Energy,
Office of Science, Office of Basic Energy Sciences under Award Number de-sc0017861. This work used the high performance computational resources provided by the Louisiana Optical Network Initiative
(http://www.loni.org), and HPC@LSU computing. This work also used the Extreme Science and Engineering Discovery Environment (XSEDE) through allocation DMR130036.

KMT is  partially supported by NSF DMR-1728457 and NSF OAC-1931445. HT has been supported by NSF OAC-1931367 and NSF DMR-1944974 grants. LC acknowledges the financial support by the Deutsche Forschungsgemeinschaft through TRR80 (project F6) Project number 107745057.

The analysis of the results was partially conducted at the Center for Nanophase Materials Sciences, which is a DOE Office of Science User Facility. 

A portion of this research was conducted at the Center for Nanophase Materials Sciences, which is a DOE Office of Science User Facility (TB). The manuscript has been authored by UT-Battelle, LLC under Contract No. DE-AC05-00OR22725 with the U.S. Department of Energy. The United States Government retains and the publisher, by accepting the article for publication, acknowledges that the United States Government retains a non-exclusive, paid-up, irrevocable, world-wide license to publish or reproduce the published form of this manuscript, or allow others to do so, for United States Government purposes. The Department of Energy will provide public access to these results of federally sponsored research in accordance with the DOE Public Access Plan (http://energy.gov/downloads/doe-public-access-plan).

\bibliography{master}% Produces the bibliography via BibTeX.

\end{document}